# Analysis of Pipelined KATAN Ciphers under Handle-C for FPGAs

Palwasha W. Shaikh and Issam W. Damaj
Electrical and Computer Engineering Department
American University of Kuwait Salmiya, Kuwait
{S00031562, idamaj}@auk.edu.kw

*Abstract*— Embedded Systems are everywhere from the smartphones we hold in our hands to the satellites that hover around the earth. These embedded systems are being increasingly integrated into our personal and commercial infrastructures. More than 98% of all processors are implanted and used in embedded systems rather than traditional computers. As a result, security in embedded systems now more than ever has become a major concern. Since embedded systems are designed to be low-cost, fast and real-time, it would be appropriate to use tiny, lightweight and highly secure cryptographic algorithms. KATAN and KATANTAN family of light-weight block ciphers are promising cryptographic options. In this paper, a sequential hardware design is developed under Handel-C. Taking a step further, Handel-C's parallel construct is taken advantage of to develop a parallel-pipelined hybrid implementation. Both sequential and parallel-pipelined implementations are tested under Altera Quartus to implement and analyze hardware designs in conjunction with DK Design Suite's Handel-C compiler. The developed designs are mapped to Altera's Stratix II that is one of the industry's highest bandwidth and density FPGAs. The results confirm that using Handel-C can provide faster implementations. The obtained results are promising and show better performance when compared with similar implementations—specifically the developed parallel-pipelined processor.

*Keywords—Hardware Design; High Performance Computing; Cryptography; Handel-C; Parallel Processing; Pipelining.*

## I. INTRODUCTION

Now more than ever securing information is of critical importance. Embedded System devices have increasingly infiltrated both our personal and industrial aspects of our daily lives. Design metrics like fast speeds, modifiability, low power consumption with small footprint and size are of primary importance; hence reusability of hardware cores is needed. Light-weight encryption algorithms like KATAN and KATANTAN are good candidates for encryption codes. Advancements in Field Programmable Gate Array (FPGA) technology coupled with the increasing availability of modern high-level hardware design tools like Handel-C can make the hardware design a relatively faster process by reducing the time to prototype and time to market [1,2,3,4].

Handel-C is a high-level programming language based on C programming language; It targets low-level hardware and is suitable for inexperienced designers. Eliminating the need for possessing in-depth hardware design experience, it enables using parallel logic in FPGA prototyping at minimum time cost [5]. Handel-C can be compiled to a number of design languages like EDIF, VHDL and Verilog and then synthesized to the corresponding hardware. Its compiler is run under DK Design Suite IDE that enables validating performance and improving product quality in less time and at a lower cost. To see this in full effect, a set of light-weight cryptographic algorithms are targeted in this investigation, namely, the KATAN family of block ciphers [6].

KATAN family of ciphers is of light-weight [6,7,8,9,10]; it is considered suitable for embedding in application that require compact size and efficient implementations [6,11,12,13,14,15]. The KATAN family is of two algorithms, namely, the KATAN and the KTANTAN. The KATAN algorithm has 32, 48, and 64-bit versions. The KTANTAN algorithm also takes the same 3 blocks of bits as the KATAN family but has a different key scheduler. Accordingly, we are targeting FPGAs to enable rapid prototyping and quick modifications to develop hardware designs and implementations.

FPGAs are the basis of reconfigurable systems. They provide several millions of gates, flexible programming, and compatibility with both low-level hardware description languages (HDLs) like VHDL and Verilog and high-level hardware languages like Handel-C. As a result, FPGAs are the first choice for designing and testing new hardware designs. Companies like Altera [16] and Xilinx [17] have made available high-end, faster bandwidth FPGAs like Virtex Pro and Stratix FPGAs for hardware designers like us.

Using the Altera's Stratix II FPGA, as the target hardware, we made attempts at developing encryption cores with appealing performance characteristics. The key motivating factors included highlighting the implementation specifics and studying the impact of the specific code system from the perspective of two modern hardware design tools. In addition, the effect of design choices on the various performance aspects is highlighted. Basically, the aim is to compare the design methodology and implementation of KATAN and KATANTAN parallel-pipelined encryption codes using high-level hardware design tools like Handel-C with relatively traditional hardware description languages like VHDL. We also wanted to identify the advantages and drawbacks of using a high-level design tool while observing its effect on performance.

In this paper, the design and implementation of several high-speed and parallel-pipelined hardware implementations for the KATAN family of block ciphers is presented. The development started by taking the software implementations of the light-weight encryption codes and modeling the designs using a hybrid model that combines parallelization and pipelining. The developed cores using Handel-C are then critically analyzed, evaluated, and benchmarked against similar implementations. The hardware cores are analyzed for their execution time, maximum frequency, clock cycles, throughput, speed ups and logic area. The targeted hardware system is Altera's Stratix II FPGA. Results are compared to similar implementations.

The paper is organized with Section 2 describing the targeted algorithms. Section 3 details the proposed hardware development. Section 4 presents the analysis, evaluation, and the comparisons with similar implementations from the literature, and Section 5 concludes the paper and plots future directions.

## II. THE KATAN FAMILY OF BLOCK CIPHERS

KATAN and KTANTAN is a family of hardware oriented block ciphers designed by de Canniere et al. in [7]. Both KATAN and KTANTAN have three variations each of 32-bit, 48-bit, and 64-bit block. All ciphers key length is of 80 bits and have a maximum round of 254, where the only difference between KATAN and KTANTAN is in the key schedule. The resulting ciphers are exteremly efficient in hardware, and offer a set of suitable solution for low-end devices that need encryption such as RFID tags.

### A. KATAN Family

KATAN 32 is the smallest of this family; however, other two members of the family encipher in a similar manner with slight variations. KATAN 32 plaintext and ciphertext are of 32 bits each. The plaintext is loaded into two registers L1, and L2 of lengths 13 and 19 bits respectively. Each round, L1 and L2 are shifted to the left such that the $i^{th}$ bit is shifted to position $(i + 1)$, thus the new computed bits are loaded in the LSB of L1 and L2. After 254 rounds of the cipher, the contents of the registers are then stored in the ciphertext array where bit 0 of L2 is the LSB of the ciphertext. KATAN32 uses two nonlinear function $fa(\cdot)$ and $fb(\cdot)$ in each round (See Equations (1) and (2)).

$$fa(L_1) = L_1[x_1] \oplus L_1[x_2] \oplus (L_1[x_3] \cdot L_1[x_4]) \oplus (L_1[x_5] \cdot IR) \oplus ka \quad (1)$$

$$fb(L_2) = L_2[y_1] \oplus L_2[y_2] \oplus (L_2[y_3] \cdot L_2[y_4]) \oplus (L_2[y_5] \cdot L_2[y_6]) \oplus kb \quad (2)$$

As the name suggests KATAN 48 and KATAN 64 will deal with plain texts of size 48-bit and 64-bit respectively. Further, in KATAN 48, in one round of the cipher the functions (1) and (2) are applied twice. The first (1) and (2) functions are applied, and then after the update of the registers, they are applied again, using the same subkeys; these steps can be done in parallel. In KATAN 64, each round applies Equations (1) and (2) three times with the same key bits.

### B. KATANTAN Family

While in the KATAN family, the 80-bit key is loaded into a register which is then repeatedly clocked. In the KTANTAN family of ciphers, the key is fixed, and the only so-called flexibility is provided by the freedom of choosing subkey bits. Thus, the design problem in the KTANTAN ciphers is choosing a sequence of subkeys in a secure, yet an efficient manner.

The KATAN and KATANTAN family is found to be secure against differential and linear attacks [6,18,19]. Several attacks based on Meet-in-the-Middle related concepts have been successfully applied on these ciphers [20]. They exploit the slow diffusion of the key material to the internal state throughout the rounds. Various hardware implementations using a 0.13μm CMOS library, doubling and tripling the use of multiplexers, and using a combination of gate equivalance that includes sequential and combinational logic for the KATAN family are presented in [7]. The authors presented several results for different design trade-offs. The highest reported speed is around 75 Kbps for both the KATAN and KTANTAN at a frequency of 100 MHz.

## III. THE DEVELOPMENT OF PARALLEL-PIPELINED HYBRID KATAN CIPHERS

The development starts by taking the software C++ implementation of KATAN Ciphers and implementing the sequential implementation in hardware under Handel-C. Next, the system was modelled using a hybrid model that combines flowcharts and concurrent process models (CPMs). Flowcharts helped in describing the sequential behavior of the algorithm, and the CPM revealed the parallel behavior of the algorithm as can be seen in Fig. 1 (a) and (b). Parallel designs are then captured using Handel-C's par construct under DK Design Suite. The used development methodology is informal, easy to use, clearly highlights the parallel code segments of the algorithm, and enables smooth capturing of the model under a high-level design tool like Handel-C. Next, the parallel implementations are taken and pipelined to create a Parallel-Pipelined Hybrid model with the use of macros and channels structures in Handel-C.

The encryption in this developed Parallel-Pipelined Hybrid model of KATAN ciphers initializes by loading an array of plaintexts with their respective keys also in an array. The plaintexts are loaded into the registers L1 and L2. The length of these two registers depends on the size of the plaintexts as discussed in Section 2. KATAN ciphers use two nonlinear functions of Equations (1) and (2), in each round, that are responsible mostly for moving bits around. The output of the Boolean functions is loaded to the LSB of the registers after they are shifted. 254 rounds are executed to insure sufficient mixing.

The encryption method is divided into three main pipelined stages. The first stage consists of three loops that initialize the plaintext and loads the key. Since these three loops are independent of one another they can run in parallel. A loop for key scheduling, and an outer loop for two nonlinear functions, (1) and (2), with two nested loops are part of the second main stage. It does most of the encryption.

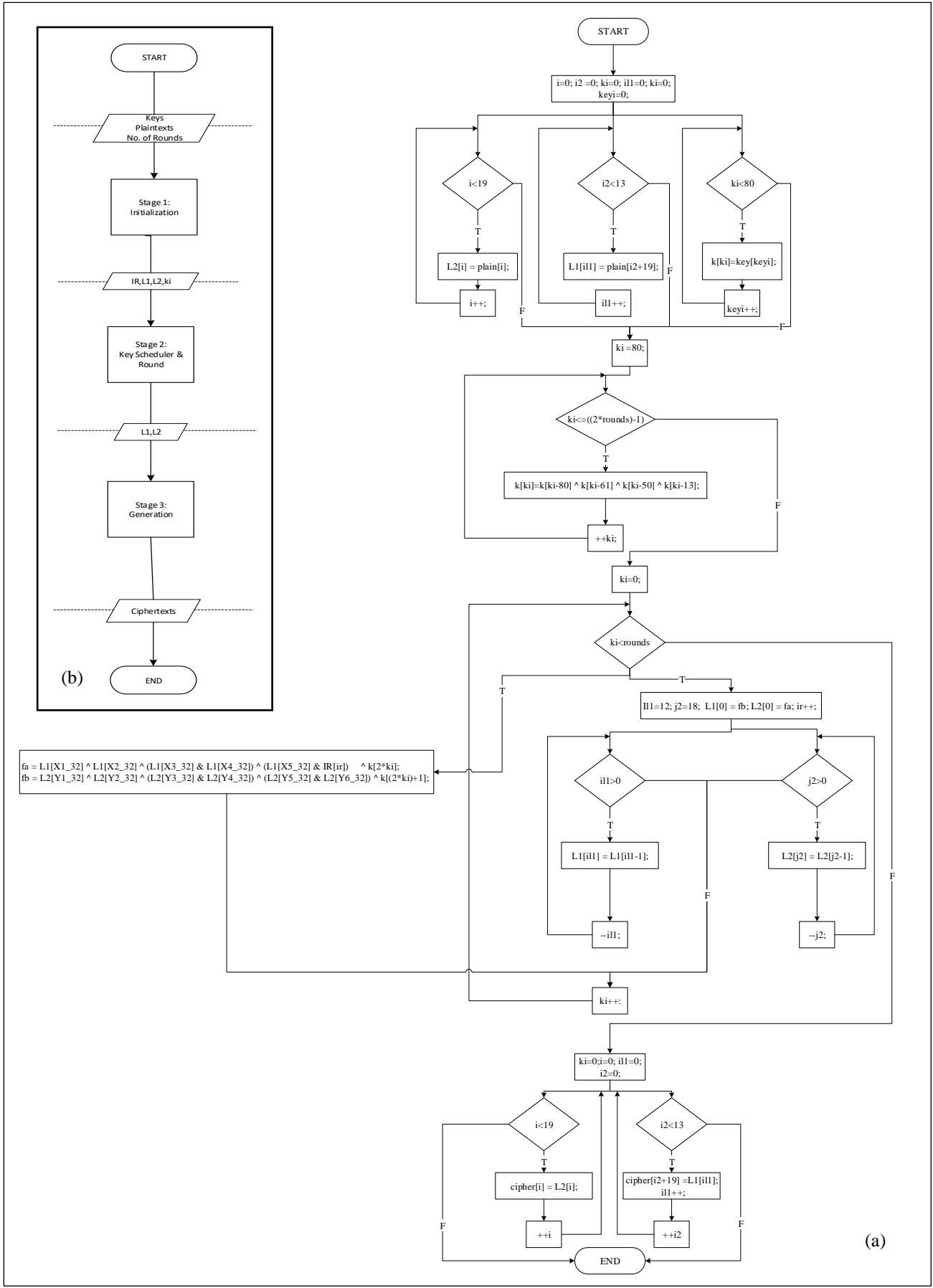

Fig. 1: (a) The hybrid model combining CPM and Flowcharts for KATAN-32. (b) The general overview of the Parallel-Pipelined Hybrid with the data transferred from one stage to other via channels construct in Handel-C.

TABLE I. HARDWARE RESULTS OF KATAN/KATANTAN BLOCK CIPHERS FOR THE SEQUENTIAL HANDEL-C IMPLEMENTATION.

| Algorithm Name | Hardware Sequential Implementation ||||||||
|---|---|---|---|---|---|---|---|---|
| | Logic Utilization (%) | Combinational ALUTs | Logic Registers | Total Clock Cycles | Fmax (MHz) | Clock Period (ns) | Total Execution Time (µs) | Throughput (Mbps) |
| KATAN 32 | 5 | 2844 | 673 | 3626 | 358.55 | 2.789 | 10.11 | 3.164 |
| KATAN 48 | 6 | 3420 | 725 | 5303 | 99.56 | 10.04 | 53.26 | 0.6008 |
| KATAN 64 | 6 | 3442 | 731 | 6928 | 94.79 | 10.55 | 73.09 | 0.4378 |
| KATANTAN 32 | 3 | 2001 | 784 | 26112 | 96.46 | 10.37 | 270.7 | 0.1182 |
| KATANTAN 48 | 3 | 2135 | 822 | 27229 | 98.75 | 10.13 | 275.7 | 0.1161 |
| KATANTAN 64 | 5 | 2080 | 841 | 28855 | 96.85 | 10.33 | 297.9 | 0.1074 |

TABLE II. HARDWARE RESULTS OF KATAN/KATANTAN BLOCK CIPHERS FOR THE PARALLEL-PIPELINED HANDEL-C IMPLEMENTATION.

| Algorithm Name | Hardware Parallel-Pipelined Hybrid Implementation |||||||||
|---|---|---|---|---|---|---|---|---|---|
| | Logic Utilization (%) | Combinational ALUTs | Logic Registers | Memory Bits | Total Clock Cycles | Fmax (MHz) | Clock Period (ns) | Total Execution Time (µs) | Throughput (Mbps) |
| KATAN 32 | 13 | 7889 | 1403 | - | 2204 | 67.22 | 14.88 | 32.7879 | 0.9759 |
| KATAN 48 | 14 | 8507 | 1852 | - | 3169 | 67.81 | 14.75 | 46.74 | 1.027 |
| KATAN 64 | 13 | 7882 | 1349 | - | 4084 | 70.02 | 14.28 | 58.33 | 1.097 |
| KATANTAN 32 | 3 | 1875 | 939 | 54 | 9703 | 110.93 | 9.015 | 87.47 | 0.3658 |
| KATANTAN 48 | 3 | 1932 | 979 | 48 | 10414 | 109.18 | 9.159 | 95.39 | 0.5032 |
| KATANTAN 64 | 2 | 1104 | 918 | 24 | 11329 | 233.81 | 4.277 | 48.45 | 1.321 |

Two loops are run in parallel to generate an array of cipher texts in the third stage. This third stage does the generation. The structure of KATAN ciphers enables the parallelization of several segments, and the overall structure is easily divided into stages for pipelining.

## IV. RESULTS AND EVALUATION

Performance analysis for the developed hardware design is done using Altera's Quartus along with DK Design Suite. The following are the definitions of the main metrics we use to analyze and evaluate the proposed developments:

- $F_{max}$ (Maximum Frequency): It indicates the clock speed that a certain core is running at.
- Number of clock cycles: The total number of clock cycles needed to finish execution.
- Total Execution Time: It is the Total number of clock cycles divided by $F_{max}$. Simply, the total time it takes a program to finish executing.
- Throughput: Number of bits encrypted over total execution time. It indicates the speed of the encryption process.
- Speed up: It is a number that measures the relative performance of two systems. The improvement in speed of execution of the same task executed on two similar processors with same architecture but different resources. Speed up in throughput is the throughput of processor 2 with respect to throughput of processor 1.
- Chip-area: The amount of logic occupied by an algorithm mapping onto an FPGA in terms of logic elements (LEs) and adaptive look-up tables (ALUTs).

Two different implementations of the KATAN ciphers under Handel-C are analyzed. First, is the sequential implementation that is similar to the original in [7]. Second, the Parallel-Pipelined Hybrid implementation. The hardware results for the sequential design are shown in Table I. Among the KATAN implementations, the 32-bit version achieved the smallest chip-area of 2844 ALUTs and 673 logic registers. On the other hand, the 64-bit version occupies the most chip-area of 3442 ALUTs and 731 logic registers. Moreover, the 32-bit version has better and faster total execution time of 10.11 µs, smallest clock period of 2.789 ns and smallest total clock cycles of 3626 with the highest operating frequency of 358.55 MHz.

Among the KATANTAN implementations, the 32-bit version has the smallest chip-area of 2001 ALUTs and 784 logic registers. Even though the 48-bit version has the highest operating frequency of 98.75 MHz, the 32-bit version has the highest throughput of 0.1182 Mbps, since the 48-bit version has higher total execution time of 275.7 µs. Overall, in the sequential implementation, KATAN 32 has the highest $F_{max}$ of 358.55 MHz and the highest throughput of 3.164 Mbps compared to all other KATAN and KATANTAN block ciphers. Whereas, the KATANTAN 32 has the smallest chip area of 2001 ALUTs and 784 registers.

Table II shows that hardware results for the parallel-pipelined hybrid. From the set of KATAN block ciphers, the 64-bit version occupies the least chip-area of 7882 ALUTs and 1349 logic registers and has the highest throughput of 1.097 Mbps with the smallest clock period of 14.28 ns and highest operating frequency of 70.02 MHz. Furthermore, the KATANTAN family's 64-bit version has the smallest chip-area of 1104 ALUTs, 918 logic registers and 24 memory bits. It has the highest operating frequency of 233.81 MHz and the smallest clock period of 4.277 ns with a throughput of 1.321 Mbps. In comparison, the KATANTAN 64 has the smallest chip area and the highest operating frequency and throughput among all KATAN and KATANTAN family of block ciphers.

TABLE III. COMPARISON OF THE SEQUENTIAL AND PARALLEL-PIPELINED IMPLEMENTATIONS; SPEED UP IS PARALLEL PIPELINED OVER THE SEQUENTIAL THROUGHPUT

| Algorithm Name | Throughput (Mbps) | | Speed Up |
|---|---|---|---|
| | Sequential | Parallel-Pipelined | |
| KATAN 32 | 3.164 | 0.9759 | 0.3084 |
| KATAN 48 | 0.6008 | 1.027 | 1.7094 |
| KATAN 64 | 0.4378 | 1.097 | 2.5643 |
| KATANTAN 32 | 0.1182 | 0.3658 | 3.0948 |
| KATANTAN 48 | 0.1161 | 0.5032 | 4.3342 |
| KATANTAN 64 | 0.1074 | 1.321 | 12.300 |

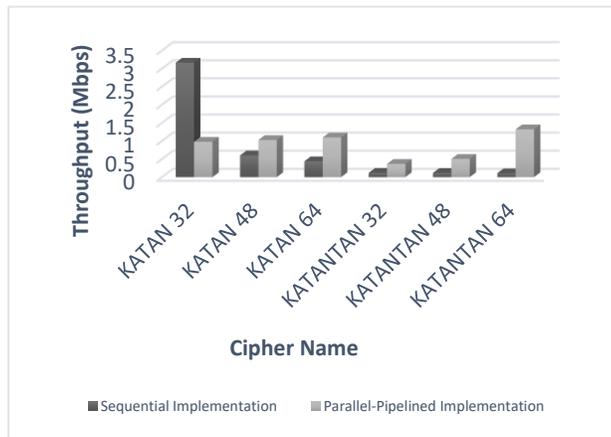

Fig. 2: Throughput of our sequential vs. parallel-pipeline implementation.

TABLE III clearly shows that the parallel-pipelined implementations in comparison to our sequential implementations has led to significant speedups especially for the KATANTAN 64-bit version with a speed up of 12.300. However, the KATAN 32-bit version is an anomaly in our speed up pattern, as the sequential implementation has higher thoroughput compared to the parallel-pipelined version as can be seen in Fig. 2. Next, the parallel-pipelined hybrid implementations are compared to the original [7] as well as the behavioral and pipelined implementations of Qatan et al. [21].

Comparing the original [7] implementation with our sequential version revealed significant speed ups as can be seen in Table IV. The parallel-pipelined implementation speed ups are bigger than sequential implementation speedups, but the 32-bit KATAN is an anomaly to the pattern with the largest speed up of 263.67 as seen in Fig. 3.

TABLE IV. COMPARISON OF THE SEQUENTIAL, PARALLEL-PIPELINED AND ORIGINAL IMPLEMENTATIONS, SPEED UP 1 IS SEQUENTIAL OVER ORIGINAL THROUGHPUT, AND SPEED UP 2 IS PARALLEL-PIPELINED OVER ORIGINAL THROUGHPUT

| Algorithm Name | Throughput (Mbps) | | | Speed $up_1$ | Speed $up_2$ |
|---|---|---|---|---|---|
| | Seq. | Parallel-Pipelined | Original [7] | | |
| KATAN 32 | 3.164 | 0.9759 | 0.012 | 263.67 | 81.33 |
| KATAN 48 | 0.6008 | 1.027 | 0.018 | 33.38 | 57.06 |
| KATAN 64 | 0.4378 | 1.097 | 0.025 | 17.51 | 43.88 |
| KATANTAN 32 | 0.1182 | 0.3658 | 0.012 | 9.85 | 30.48 |
| KATANTAN 48 | 0.1161 | 0.5032 | 0.018 | 6.45 | 28.00 |
| KATANTAN 64 | 0.1074 | 1.321 | 0.025 | 4.30 | 52.84 |

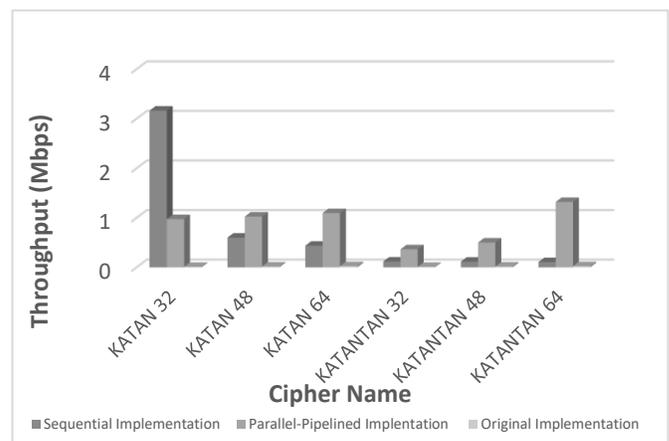

Fig. 3: Throughput of our sequential vs. our parallel-pipeline implementation vs. the original implementation.

TABLE V. COMPARISON OF THE SEQUENTIAL, PARALLEL-PIPELINED AND BEHAVIORAL [21] IMPLEMENTATIONS, SPEED UP 1 IS SEQUENTIAL OVER BEHAVIORAL THROUGHPUT, AND SPEED UP 2 IS PARALLEL-PIPELINED OVER BEHAVIORAL THROUGHPUT

| Algorithm Name | Throughput (Mbps) | | | Speed $up_1$ | Speed $up_2$ |
|---|---|---|---|---|---|
| | Seq. | Parallel-Pipelined | Behav. [21] | | |
| KATAN 32 | 3.164 | 0.9759 | 21.76 | 0.145 | 0.045 |
| KATAN 48 | 0.6008 | 1.027 | 25.39 | 0.024 | 0.041 |
| KATAN 64 | 0.4378 | 1.097 | 26.89 | 0.016 | 0.041 |
| KATANTAN 32 | 0.1182 | 0.3658 | - | - | - |
| KATANTAN 48 | 0.1161 | 0.5032 | - | - | - |
| KATANTAN 64 | 0.1074 | 1.321 | - | - | - |

As can be seen in Table V and Table VI, the behavioral design and pipeline designs from [21] respectively are significantly faster in comparison to our design. This was expected as our designs have smaller clock periods but larger number of clock cycles whereas both the behavioral and pipeline implementations in [21] have smaller number of clock cycles but larger clock periods. In other words, comparing our designs with [21] reveals the flaw in high-level design tools like Handel-C. Even though Handel-C allows for faster implementation and saves time, but tradeoffs in terms of control over the clock cycle and clock period must be made.

TABLE VI. COMPARISON OF THE SEQUENTIAL, PARALLEL-PIPELINED AND PIPELINE [21] IMPLEMENTATIONS, SPEED UP 1 IS SEQUENTIAL OVER PIPELINE THROUGHPUT, AND SPEED UP 2 IS PARALLEL-PIPELINED OVER PIPELINE THROUGHPUT

| Algorithm Name | Throughput (Mbps) | | | Speed up$_1$ | Speed up$_2$ |
|---|---|---|---|---|---|
| | Seq. | Parallel-Pipelined | Pipeline [21] | | |
| KATAN 32 | 3.164 | 0.9759 | 312.19 | 0.01 | 0.0031 |
| KATAN 48 | 0.6008 | 1.027 | 355.55 | 0.00169 | 0.0030 |
| KATAN 64 | 0.4378 | 1.097 | 426.66 | 0.00088 | 0.0026 |
| KATANTAN 32 | 0.1182 | 0.3658 | - | - | - |
| KATANTAN 48 | 0.1161 | 0.5032 | - | - | - |
| KATANTAN 64 | 0.1074 | 1.321 | - | - | - |

## V. CONCLUSION

This paper presents the hardware implementations of the KATAN and KATANTAN family of block ciphers using a high-level design tool like Handel-C. We first implemented a sequential hardware design and then progressed to a parallel-pipelined hybrid implementation. These designs were then taken and analyzed on a high-performance Altera Stratix II FPGA. Using the analysis tools, it was revealed that both our sequential and parallel-pipelined hybrid are faster than the original implementations in [7]. The KATAN 32's sequential implementation has the highest speed-up of 263.67 compared to the original implementation. The latter case being the exception, the parallel-pipelined implementations were generally faster with KATAN 32 having the highest speed up of 81.33. Importantly, advances in hardware programming languages has led to the creation of a high-level hardware design tool like Handel-C, there are some significant tradeoffs for the reduction in time to prototype and time to market.

These drawbacks include a significant impact on performance. Designers will have to surrender control over the clock cycles and clock period that low level hardware language like VHDL allows for in [21]. For instance, the VHDL implementation of [21] takes 3 clock cycles with a very low frequency of 24.190 MHz for the KATAN 32 pipeline implementation compared to ours parallel-pipelined hybrid's 67.22 MHz with 2204 clock cycles. Good designs require good compromises; without a doubt, Handel-C allows for saving time, but the careful control on clock period should be reconsidered.

It could be concluded that DK Design Suite is a relatively easy and simple tool. With a relatively quick learning curve, it allows users to write programs in Handel-C. Since Handel-C has elements of C programming language, it makes it a great choice for inexperienced hardware designers. Additionally, for hardware, Handel-C enables parallelism, bit manipulation, channels, macros, etc. It is a versatile tool for hardware programming that can opens gates to more designers to innovate. Also, it can pave a path for similar or better programming languages to be developed. Future works include critiquing and highlighting the true potential and drawbacks of high-level hardware design tools.